\documentclass[11pt,showkeys,superscriptaddress]{revtex4-2}
\usepackage{amsmath}
\usepackage{amssymb}
\usepackage{graphicx}
\usepackage{ulem,xcolor}
\usepackage{slashed}

\DeclareMathOperator{\sech}{sech}

\begin{document}
\raggedbottom
	
\title{
	Spinor domain wall and test fermions on an arbitrary domain wall
}
	
\author{
	Vladimir Dzhunushaliev
}
\email{v.dzhunushaliev@gmail.com}
	
\affiliation{
	Department of Theoretical and Nuclear Physics,  Al-Farabi Kazakh National University, Almaty 050040, Kazakhstan
}

\affiliation{
	Institute of Experimental and Theoretical Physics,  Al-Farabi Kazakh National University, Almaty 050040, Kazakhstan
}

\affiliation{
	Academician J.~Jeenbaev Institute of Physics of the NAS of the Kyrgyz Republic, \\265a, Chui Street, Bishkek 720071, Kyrgyzstan
}

\affiliation{
Laboratory for Theoretical Cosmology, International Centre of Gravity and Cosmos,
Tomsk State University of Control Systems and Radioelectronics (TUSUR),
Tomsk 634050, Russia
}
	
\author{Vladimir Folomeev}
\email{vfolomeev@mail.ru}

\affiliation{
	Institute of Experimental and Theoretical Physics,  Al-Farabi Kazakh National University, Almaty 050040, Kazakhstan
}

\affiliation{
	Academician J.~Jeenbaev Institute of Physics of the NAS of the Kyrgyz Republic, \\265a, Chui Street, Bishkek 720071, Kyrgyzstan
}

\affiliation{
Laboratory for Theoretical Cosmology, International Centre of Gravity and Cosmos,
Tomsk State University of Control Systems and Radioelectronics (TUSUR),
Tomsk 634050, Russia
}

\author{
	Dina Zholdakhmet
}
\email{dinaicloud97@gmail.com}
	
\affiliation{
	Department of Theoretical and Nuclear Physics,  Al-Farabi Kazakh National University, Almaty 050040, Kazakhstan
}

\affiliation{
	Institute of Experimental and Theoretical Physics,  Al-Farabi Kazakh National University, Almaty 050040, Kazakhstan
}
	
\date{\today}
	
\begin{abstract}
We consider a spinor domain wall embedded in a five-dimensional spacetime with a nondiagonal metric.
The corresponding plane symmetric solutions for linear and nonlinear spinor fields with different parameters are obtained.
It is shown that in the general case the metric functions and spinor fields do not possess $Z_2$ symmetry with respect to the domain wall.
We study the angular momentum density of the domain wall arising because of the presence of the spinor field creating the wall.
The properties of test fermions located on an arbitrary domain wall are considered.
The concepts of the ``second spin'' (arising due to the properties of the Lorentz group generators in a five-dimensional spacetime)
and of the ``second magnetic field'' (representing the components $F_{i 5}$ of the electromagnetic field five-tensor) are introduced.
We find eigenspinors of the ``second spin'' and show that some of them represent the Bell states.
In the nonrelativistic limit we derive the Pauli equation for the test fermions on the domain wall which contains an extra term
describing the interaction of a spin-$1/2$ particle with the ``second magnetic field''; this allows the possibility of an experimental verification of the existence of extra dimensions.
\end{abstract}
	
	
\keywords{
Thick domain wall  solutions, Dirac equation, ``second spin'', ``second magnetic field'', test fermions, Pauli equation, experimental verification
}
	
\date{\today}
	
\maketitle
	
\section{Introduction}

The study of self-consistent solutions to the Einstein-Dirac equations is a fascinating and very difficult problem. Such solutions are few at present.
The main difficulty is that spinors in the Dirac equation have a spin and hence there are no spherically symmetric solutions:
they must be at least axially symmetric, which greatly complicates deriving solutions describing a gravitating spinor field.
To simplify matters, one can take two spinor fields with oppositely directed spins, as is done,  for example, in Refs.~\cite{Finster:1998ws,Finster:1998ux}
for particlelike solutions. In cosmology, classical spinor fields have been considered in Refs.~\cite{Saha:2003xv,Saha:2005ya,Saha:2006iu,Saha:2010zza}.
In astrophysics,  stars supported by spinor fields have been studied
in Refs.~\cite{Herdeiro:2017fhv,Dzhunushaliev:2018jhj,Dzhunushaliev:2019kiy,Herdeiro:2019mbz,Dzhunushaliev:2019uft,Herdeiro:2021jgc}.
[Note here that Refs.~\cite{Herdeiro:2019mbz,Herdeiro:2021jgc} deal with a single spinor field describing spinning (axially symmetric) configurations.]

A consideration of a hypothesis, called brane world scenario, according to which we live on a thin leaf (brane) embedded into some multidimensional
space (bulk) is of special interest. In constructing brane models, one usually either uses various scalar fields
(for a review, see Ref.~\cite{Dzhunushaliev:2009va}) or performs modeling within some extended theories of gravity (see the review~\cite{Liu:2017gcn}).
However, it is of some interest to employ another fundamental fields as a matter source supporting the brane.
In this connection, in the present paper we continue our previous investigations began  in Refs.~\cite{Dzhunushaliev:2010fqo,Dzhunushaliev:2011mm}
where  thick brane solutions supported by nonlinear spinor fields have been considered (see also the recent work~\cite{Cui:2022djf}
where  five-dimensional domain wall  solutions supported by a nonlinear spinor field are under investigation).
Working within the Einstein-Dirac theory, in the first part of the current paper, we study five-dimensional
plane symmetric solutions describing domain walls. The main distinctive feature of the present study is the use of
a nondiagonal metric; this logically follows from the fact that a spinor field must possess a spin whose presence
will in general result in the appearance of a nondiagonal component in the metric which describes a rotation.
We will examine the case where  gravitation interacts both with linear and nonlinear spinor fields.
The nonlinear spinor field is of special interest, since in the absence of gravity the nonlinear spinor field (unlike the linear one)
permits solitonlike solutions (see, e.g., Refs~\cite{Finkelstein:1951zz,Finkelstein:1956}).

In the second part of the paper, we study the properties and behavior of \textit{test} fermions living on
 \textit{any} domain wall. It will be shown that the spatial part of the five-dimensional spin tensor splits into an ordinary spin and a ``second spin''
 which is represented by the components  $\Sigma_{5 i}$ of the five-dimensional spin tensor. In order to study the characteristics of the motion
 of the test fermions on the domain wall, we will obtain a nonrelativistic  approximation for the five-dimensional Dirac equation (the Pauli equation)
 which will contain extra terms (compared with the Pauli equation in our four-dimensional spacetime). One of these terms involves the
 component $A_5$ of the five-potential of an electromagnetic field, and another one contains the direct interaction between a spin and
a ``second magnetic field'' which is introduced as a vector $\mathcal{H}_i = F_{i 5}$, where $F_{A B}$ is the electromagnetic field five-tensor.
It is remarkable that the presence of such interaction may lead to experimentally measurable consequences;
this in turn may permit one to verify the hypothesis that our world is a domain wall (or a brane) embedded in some multidimensional space (bulk).

In the present study we regard spinor fields as classical ones, although it is known that in nature they are described as quantum fields. 
In this connection, the question arises as to when one can systematically regard fermionic quantum fields as classical spinors?
This question is considered, in particular, in Ref.~\cite{Armendariz-Picon:2003wfx} where the conclusion has been drawn that a classical spinor field may appear either as a result of
some effective description of a more complicated quantum system or when a quantum state of a spinor is in some sense ``close'' to a vacuum state
where a classical consideration of a massive Dirac spinor may be a good approximation.

It is worth mentioning that, strictly speaking, the configurations studied here cannot yet be referred to as  branes, since this
would demand to demonstrate that the domain walls under consideration are able to trap
 zero modes of various matter fields corresponding to particles or fields in the Standard Model.
In particular, for fermions, this can be done, for example, analogously to the problem of trapping particles with spin $1/2$ on the domain wall
considered in the pioneering paper~\cite{Rubakov:1983bb} where the authors studied  the question of trapping massless
fermions on the domain wall supported by a scalar field. Later on a number of works have been done devoted to the question of localization of 
Dirac fermion fields on thick branes (for a review, see, e.g., Ref.~\cite{Liu:2017gcn})
 For our case, this question must be considered separately.

\section{Field equations}
\label{gen_equations}

We work within the five-dimensional theory of gravitation with a source of matter in the form of a nonlinear spinor field $\psi$  with the total action (hereafter we work in  units $8\pi G=c=\hbar=1$) 
\begin{equation*}
	S = \int \left( - \frac{R}{2} + \Lambda + \mathcal L_m  \right) \sqrt{g} \; d^5 x ,
\end{equation*}
where $R$ is the five-dimensional scalar curvature, $\Lambda$ is the five-dimensional cosmological constant,  
$g$ is the determinant of the five-dimensional metric  $g_{A B}$ with $A, B = 0,1,2,3,5$ being the world index. 
For our purposes, we have taken the plus sign in front of  the cosmological constant $\Lambda$. In turn, the Lagrangian of the nonlinear spinor field is of the form
\begin{equation*}
 \mathcal L_m = \frac{\imath}{2} \left(
		\bar \psi \slashed{\nabla} \psi - \bar \psi \overleftarrow{\slashed{\nabla}} \psi
	\right) - m \bar \psi \psi  + V(\bar \psi, \psi),
\end{equation*}
where $m$ is some parameter and the potential $V(\bar \psi, \psi)$ is understood to be so chosen that the condition
$
    \bar \psi \frac{\partial V}{\partial \bar \psi} = 2 V
$
holds.
In what follows, we will use the potential
$$
V=\frac{\lambda}{2}\left(\bar\psi \psi\right)^2 ,
$$
where $ \lambda$ is some parameter.
The corresponding five-dimensional Einstein and Dirac equations are
\begin{eqnarray}
E_{ab}\equiv  R_{ab} - \frac{1}{2} \eta_{ab} R + \eta_{ab} \Lambda- T_{ab}	&=&0,
\label{2-20} \\
	\left(
	\imath \Gamma^a e_a^{\phantom{a} A}  D_A - m +
	\frac{\partial V}{\partial \bar \psi}
	\right) \psi &=& 0,
\label{2-30}
\end{eqnarray}
where $a,b=\bar 0, \bar 1, \bar 2, \bar 3, \bar 5$ are the Lorentz indices;
$A=0,1,2,3,5$ is the world index; $e^a_{\phantom{a}A}$ is the 5-bein; $\Gamma^a$
are the five-dimensional Dirac matrices in flat Minkowski space;
$D_A \psi = \left( \partial_A - \frac{1}{4} \omega_A^{\phantom{A} ab}
\Gamma_{ab} \right) \psi$ is the covariant derivative of the spinor $ \psi$;
$\Gamma_{ab} = \frac{1}{2}\left(\Gamma_a\Gamma_b-\Gamma_b\Gamma_a\right)$;
$\slashed{\nabla} \psi=e^A_a \gamma^a D_A \psi$;
  $\eta_{ab}=\text{diag}\left(1,-1,-1,-1,-1\right)$ is the five-dimensional
covariant Minkowski metric. According to the textbook \cite{ortin}, the energy-momentum tensor for the spinor field is taken in the form
\begin{equation}
 T_a^{\;A } = \frac{\imath}{2} \bar \psi \left(
		\Gamma^A e_a^{\phantom{a} B} + \Gamma_a g^{AB}
	\right) D_B \psi -
    \frac{\imath}{2} D_B \bar \psi \left(
		\Gamma^A e_a^{\phantom{a} B} + \Gamma_a g^{AB}
	\right) \psi -
    e_a^{\phantom{a} A} \mathcal L_m,
\label{2-40}
\end{equation}
where $\Gamma^A = e_a^{\,\, A} \Gamma^a$ are the five-dimensional Dirac matrices in a curved spacetime;
$g^{AB} = e_a^{\phantom{a} A} e_b^{\,\, B} \eta^{ab}$ is the five-dimensional contravariant metric tensor;
$\bar \psi =  \psi^\dagger \Gamma^{\bar 0}$ is the Dirac conjugate spinor;
$
	D_A \bar \psi = \bar \psi\left ( \overleftarrow \partial_A + \frac{1}{4}
	\omega_A^{\phantom{A} ab} \Gamma_{ab} \right)
$ with
$\bar \psi \overleftarrow \partial_A = \partial_A \bar \psi$. Notice here that our definition of the energy-momentum tensor \eqref{2-40}
has the opposite sign compared with Ref.~\cite{ortin} in order to be consistent with the definitions for $R_{ab}$ from Ref.~\cite{poplawski}.

The five-dimensional Dirac matrices in flat Minkowski space are
$$
  \Gamma^{\bar 0} = \begin{pmatrix}
		0												&		\mathbb I_{2 \times 2} \\
		\mathbb I_{2 \times 2} 	& 0
	\end{pmatrix}, \quad
  \Gamma^{\bar i} = \begin{pmatrix}
		0					&		-\sigma_{\bar i} \\
		\sigma_{\bar i}  & 0
	\end{pmatrix} \,\text{with}\,\, \bar i = 1,2,3,
 \quad
  \Gamma^{\bar 5} = \begin{pmatrix}
		-\imath \mathbb I_{2 \times 2}	&		0 \\
		0   											& \imath \mathbb I_{2 \times 2}
	\end{pmatrix},
$$
where $\mathbb I_{2 \times 2}$ is the $2 \times 2$ unit matrix, and $\sigma_{\bar i}$
are the Pauli matrices
$$
  \sigma_{\bar 1} = \begin{pmatrix}
		0	 &		1 \\
		1  & 0
	\end{pmatrix}, \quad
\sigma_{\bar 2} = \begin{pmatrix}
		0	 & -\imath \\
		\imath  & 0
	\end{pmatrix}, \quad
\sigma_{\bar 3} = \begin{pmatrix}
		1	 &	0 \\
		0  & -1
	\end{pmatrix}.
$$
We seek a wall-like solution of the system \eqref{2-20} and \eqref{2-30}. To do this, let us choose the following orthonormal 5-bein:
\begin{equation}
e^0_A d x^A=\chi(r) d t, \quad e^1_A d x^A=\phi(r) d x, \quad e^2_A d x^A=\phi(r) d y, \quad e^3_A d x^A=\xi(r) d t+\eta(r) d z, \quad
e^5_A d x^A= d r,
\label{5-bein}
\end{equation}
such that 
\begin{equation}
ds^2\equiv\eta_{a b}\left(e^a_A d x^A\right)\left(e^b_B d x^B\right)=\left(\chi^2-\xi^2\right)d t^2-\phi^2\left(dx^2+dy^2\right)-\eta^2 dz^2-dr^2-2\,\xi\eta\, dt\, dz .
\label{metr}
\end{equation}

In turn, for the spinor field, we employ the $\mathfrak{Ansatz}$
\begin{equation}
 \psi = e^{\imath (\Omega t+ M z)} \begin{pmatrix}
		A(r) 	\\
		0 		\\
		B(r)	\\
		0
	\end{pmatrix},
\label{2-110}
\end{equation}
where $\Omega$ and $M$ are some constants. Substituting this $\mathfrak{Ansatz}$ and
5-bein from \eqref{5-bein} in Eqs.~\eqref{2-20} and \eqref{2-30} and taking into account \eqref{2-40}, one can obtain
the following set of the Einstein-Dirac equations:
\begin{eqnarray}
\label{En1}
	\frac{\xi^{\prime\prime}}{\xi} &=& -\frac{5}{12}\frac{\xi^{\prime 2}}{\chi^2}-\frac{5}{12}\frac{\xi^2}{\chi^2}\frac{\eta^{\prime 2}}{\eta^2}
	+\frac{\eta^{\prime 2}}{\eta^2}+\frac{1}{3}\frac{\phi^{\prime 2}}{\phi^2}+\frac{2}{3}\frac{\eta^{\prime }}{\eta}\frac{\phi^{\prime }}{\phi}
	-\frac{5}{3}\frac{\eta^{\prime }}{\eta}\frac{\chi^{\prime }}{\chi}+\frac{2}{3}\frac{\phi^{\prime }}{\phi}\frac{\chi^{\prime }}{\chi}
	+\frac{5}{6}\frac{\xi\eta^\prime\xi^\prime}{\eta \chi^2}-\frac{\eta^{\prime }}{\eta}\frac{\xi^{\prime }}{\xi}
	-2\frac{\xi^{\prime }}{\xi}\frac{\phi^{\prime }}{\phi}+\frac{\xi^{\prime }}{\xi}\frac{\chi^{\prime }}{\chi}
\nonumber \\
	&-&\frac{1}{2}\frac{ A B\left[\xi^2\eta^\prime-\eta\xi\xi^\prime+2\chi\left(\eta\chi^\prime-\chi\eta^\prime\right)\right]}{\eta\xi\chi}
	+\frac{1}{3}\Big\{\frac{A^2\left[M\left(3\chi^2+\chi\xi-\xi^2\right)+\eta\left(\xi\left(\Omega-2\lambda\chi B^2\right)-3\Omega\chi\right)\right]}{\eta\xi\chi}
\nonumber\\
	&+&\frac{B^2\left[\Omega\eta\left(\xi+3\chi\right)-M\left(\xi^2+\xi\chi-3\chi^2\right)\right]}{\eta\xi\chi}+\Lambda
	\Big\} ,
 \\
\label{En2}
	\frac{\chi^{\prime\prime}}{\chi} &=& \frac{7}{12}\frac{\xi^{\prime 2}}{\chi^2}+\frac{7}{12}\frac{\xi^2}{\chi^2}\frac{\eta^{\prime 2}}{\eta^2}+
	\frac{1}{3}\frac{\phi^{\prime 2}}{\phi^2}+\frac{2}{3}\frac{\eta^{\prime }}{\eta}\frac{\phi^{\prime }}{\phi}-
	\frac{2}{3}\frac{\eta^{\prime }}{\eta}\frac{\chi^{\prime }}{\chi}-\frac{4}{3}\frac{\phi^{\prime }}{\phi}\frac{\chi^{\prime }}{\chi}
	-\frac{7}{6}\frac{\xi\eta^\prime\xi^\prime}{\eta \chi^2}
\nonumber\\
	&-&
	\frac{1}{2} \frac{ A B\left(\eta\xi^\prime-\xi\eta^\prime\right)}{\eta\chi}
	-	\frac{1}{3}
	\left\lbrace
		\frac{A^2 \left[
			-M \left(2\xi + \chi\right) + 2\eta\left(\Omega + \lambda \chi B^2\right)
			\right]}{\eta\chi}
		+	\frac{B^2\left[2\Omega\eta
		+ M\left(\chi-2\xi\right)\right]}{\eta\chi}
	- \Lambda
	\right\rbrace  ,
\\
\label{En3}
	\frac{\eta^{\prime\prime}}{\eta} & = & - \frac{5}{12}
	\frac{\xi^{\prime 2}}{\chi^2}
	- \frac{5}{12}\frac{\xi^2}{\chi^2}\frac{\eta^{\prime 2}}{\eta^2}
	+ \frac{1}{3}\frac{\phi^{\prime 2}}{\phi^2}
	- \frac{4}{3}\frac{\eta^{\prime }}{\eta}\frac{\phi^{\prime }}{\phi}
	-	\frac{2}{3}\left(\frac{\eta^{\prime }}{\eta}\frac{\chi^{\prime }}{\chi}
	- \frac{\phi^{\prime }}{\phi}\frac{\chi^{\prime }}{\chi}\right)
	+ \frac{5}{6}\frac{\xi\eta^\prime\xi^\prime}{\eta \chi^2}
\nonumber\\
	& - &
	\frac{1}{2}\frac{ A B\left(\xi\eta^\prime - \eta\xi^\prime\right)}{\eta\chi} -
	\frac{1}{3}
	\left\lbrace
		\frac{A^2\left[
				M\left(\xi + 2\chi\right) - \eta\left(\Omega - 2\lambda \chi B^2\right)
			\right]}{\eta\chi} -
		\frac{B^2 \left[\Omega\eta+M\left(2\chi-\xi\right)\right]}{\eta\chi} - \Lambda
	\right\rbrace ,
\\
\label{En4}
	\frac{\phi^{\prime\prime}}{\phi} &=& \frac{1}{12}\frac{\xi^{\prime 2}}{\chi^2}
	+ \frac{1}{12}\frac{\xi^2}{\chi^2}\frac{\eta^{\prime 2}}{\eta^2}
	- \frac{2}{3}\frac{\phi^{\prime 2}}{\phi^2}
	- \frac{1}{3}
	\left(
		\frac{\eta^{\prime }}{\eta}\frac{\phi^{\prime }}{\phi}
		- \frac{\eta^{\prime }}{\eta}\frac{\chi^{\prime }}{\chi}
		+ \frac{\phi^{\prime }}{\phi}\frac{\chi^{\prime }}{\chi}
	\right) - \frac{1}{6}\frac{\xi\eta^\prime\xi^\prime}{\eta \chi^2}
\nonumber \\
	&-&
	\frac{1}{3} \left\lbrace
		\frac{B^2\left[M\left(\xi+\chi\right)-\Omega\eta\right]}{\eta\chi}-\frac{A^2\left[M\left(\chi-\xi\right)+
		\eta\left(\Omega-2\lambda\chi B^2\right)
		\right]}{\eta\chi}-\Lambda
	\right\rbrace ,
\\
\label{DE1}
	A^\prime &= & -A\left[
		\frac{\left(\xi+2\chi\right)\eta^\prime}{4\eta\chi}
		- \frac{\xi^\prime}{4\chi}
		+ \frac{\phi^\prime}{\phi}
		+ \frac{\chi^\prime}{2\chi} - m
	\right] - 2 \lambda A^2 B
	+ \Omega\frac{ B}{\chi}
	- M\frac{ B\left(\xi+\chi\right)}{\eta\chi} ,
\\
\label{DE2}
	B^\prime & = & B\left[
		\frac{\left(\xi - 2\chi\right)\eta^\prime}{4\eta\chi}
		- \frac{\xi^\prime}{4\chi}
		- \frac{\phi^\prime}{\phi}
		- \frac{\chi^\prime}{2\chi} - m
		\right] + 2 \lambda A B^2
		- \Omega\frac{ A}{\chi}
		+ M \frac{ A\left(\xi-\chi\right)}{\eta\chi},
\end{eqnarray}
where the prime denotes differentiation with respect to the fifth coordinate $r$. The above gravitational equations represent respectively the following combinations
of the Einstein equations~\eqref{2-20}:
$$
-\frac{1}{3}E_{\bar{0}\bar{0}}+\frac{2}{3}E_{\bar{1}\bar{1}}-\frac{2}{3}E_{\bar{3}\bar{3}}+E_{\bar{0}\bar{3}}=0,\,
\frac{2}{3}E_{\bar{0}\bar{0}}+\frac{2}{3}E_{\bar{1}\bar{1}}+\frac{1}{3}E_{\bar{3}\bar{3}}=0,\,
-\frac{1}{3}E_{\bar{0}\bar{0}}+\frac{2}{3}E_{\bar{1}\bar{1}}-\frac{2}{3}E_{\bar{3}\bar{3}}=0,\,
-E_{\bar{0}\bar{0}}-E_{\bar{1}\bar{1}}+E_{\bar{3}\bar{3}}=0 .
$$
This enabled us to write them in the form where each equation contains a higher derivative only of one metric function.
In turn, in addition to these gravitational equations, we also have the constraint equation [$(\bar{5}-\bar{5})$-component of the Einstein equations~\eqref{2-20}]
\begin{eqnarray}
\label{En_constr}
&&\frac{\xi^2\eta^{\prime 2}}{4\eta^2\chi^2}-\frac{\xi\eta^\prime\xi^\prime}{2\eta\chi^2}+\frac{\xi^{\prime 2}}{4\chi^2}+
\frac{2\eta^\prime\phi^\prime}{\eta\phi}+\frac{\phi^{\prime 2}}{\phi^2}+\frac{\eta^\prime\chi^\prime}{\eta\chi}+2\frac{\phi^\prime\chi^\prime}{\phi\chi}
\nonumber\\
&&=\Lambda-2 m A B+2\lambda A^2 B^2-\Omega\frac{A^2+B^2}{\chi}+M\frac{A^2\left(\xi-\chi\right)+B^2\left(\xi+\chi\right)}{\eta\chi},
\end{eqnarray}
which contains only first derivatives of the metric functions. It will be used below in assigning boundary conditions.

\section{Domain wall solutions}

The equations \eqref{En1}-\eqref{DE2} permit a number of solutions given below. To derive these solutions, we will begin from the fact that in the neighbourhood of the domain wall $r=0$
the solutions can be represented as a power series in $r$,
\begin{eqnarray}
\label{BC_metr}
&&\xi\approx \xi_0+\frac{\xi_2}{2}r^2 ,\quad
\chi\approx \chi_0+\frac{\chi_2}{2}r^2 ,\quad
\eta\approx \eta_0+\frac{\eta_2}{2}r^2 ,\quad
\phi\approx \phi_0+\frac{\phi_2}{2}r^2 ,\\
\label{BC_spinor}
&&A\approx A_0+A_1 r ,\quad
B\approx B_0+B_1 r .
\end{eqnarray}
These expansions will be used as boundary conditions in solving the equations~\eqref{En1}-\eqref{DE2}.

\subsection{The case of $\Omega=M=0$}
\label{Omega_M_0}

\begin{figure}[t]
	\includegraphics[width=1\linewidth]{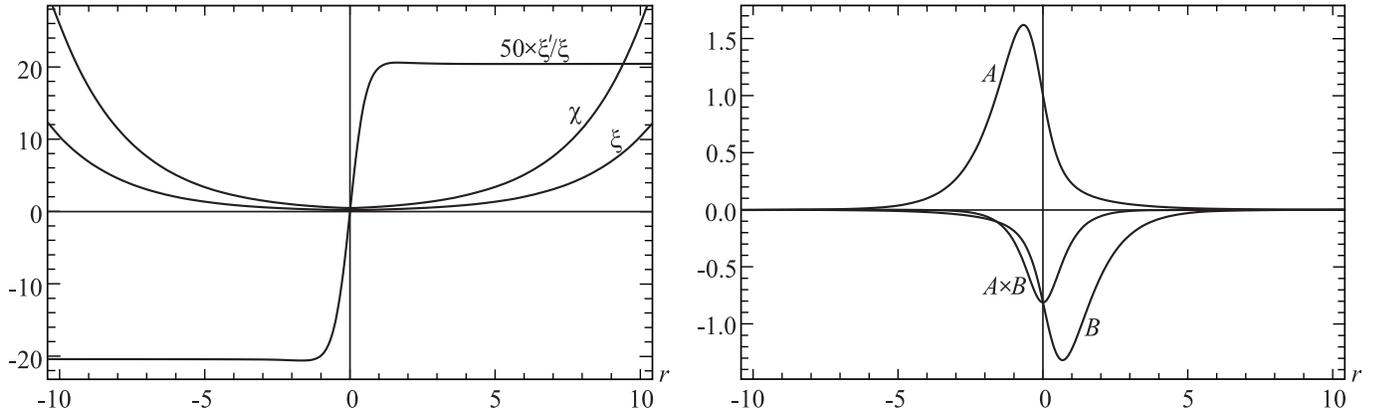}
	\caption{Examples of the solutions with $\Omega=M=0$ and with $\Lambda=1$, $\lambda=-1$, $m=0.2$.
The solutions for the metric functions $\eta$ and $\phi$ follow from Eq.~\eqref{lin_dep} for an appropriate choice of the arbitrary parameters
 $\beta$ and $\gamma$.
	}
	\label{fig_sym_sol}
\end{figure}

Consider first the simplest case of static solutions with $\Omega=M=0$. For this case, the Dirac equations~\eqref{DE1} and \eqref{DE2}
can be integrated analytically in the form
$$
A B=\frac{C}{\chi \eta \phi^2} ,
$$
where $C$ is an integration constant. In this case, it can be shown that solutions for the metric functions are linearly dependent:
\begin{equation}
\xi=\alpha \chi, \quad\eta=\beta \chi,  \quad\phi=\gamma \chi,
\label{lin_dep}
\end{equation}
where $\alpha,\beta$, and $\gamma$~are some constants, and $0<\alpha<1$ to ensure that the signature of the metric~\eqref{metr} remains unchanged.
As a result, the Einstein equations~\eqref{En1}-\eqref{En4} reduce to one equation
\begin{equation}
\frac{\chi^{\prime\prime}}{\chi}+\left(\frac{\chi^{\prime}}{\chi}\right)^2+\frac{2\tilde{C}^2\lambda}{3\chi^8}-\frac{\Lambda}{3}=0,
\label{chi_eq}
\end{equation}
and the constraint equation \eqref{En_constr} yields
\begin{equation}
6\left(\frac{\chi^{\prime}}{\chi}\right)^2+\frac{2\tilde{C} m}{\chi^4}-
\frac{2\tilde{C}^2\lambda}{\chi^8}-\Lambda=0,
\label{constr_eq}
\end{equation}
where the new constant $\tilde{C}=C/\left(\beta\gamma^2\right)$ is introduced. Taking into account the boundary conditions~\eqref{BC_metr},
one can find the following particular solution of the equation~\eqref{chi_eq} that also satisfies the
constraint equation~\eqref{constr_eq} (this is achieved by suitably adjusting the integration constant $\tilde{C}$):
$$
\chi= \left\{-\frac{\tilde{C}\, m \left[\sqrt{1-\frac{2\lambda\Lambda}{m^2}}\cosh{\left(2\sqrt{\frac{2\Lambda}{3}}r\right)}-1
\right]}{\Lambda}
\right\}^{1/4} \quad \text{with} \quad \tilde{C}=\frac{m \chi_0^4}{2\lambda}\left(1+\sqrt{1-\frac{2\lambda\Lambda}{m^2}}\right).
$$
It is evident that this solution is symmetric with respect to the domain wall located at $r=0$.
(Notice here that, using this solution, one can find the spinor functions $A$ and $B$ in an analytical form;
however, since these expressions are too cumbersome, we do not show them here.)
The corresponding solutions are exemplified in Fig.~\ref{fig_sym_sol}. It is worth noting here that, in the case under consideration, the solutions for the spinor
functions $A$ and $B$ are asymmetric, while their product
$A B\equiv \tilde{C}/\chi^4$~is symmetric with respect to the domain wall (see the right panel of Fig.~\ref{fig_sym_sol}).

Asymptotically  (as $r\to \pm\infty$), the spinor functions  $A$ and $B$ tend to zero, and the metric functions exhibit an exponential growth. Accordingly, the scalar curvature
\begin{eqnarray}
	R & = & \frac{8}{3}\Lambda
	- \frac{16}{3}\lambda A^2 B^2
	+ \frac{2}{3}\frac{\Omega}{\chi}\left(A^2+B^2\right)
	-	\frac{2}{3}\frac{M}{\eta\chi}
	\left[A^2\left(\xi-\chi\right) + B^2\left(\xi+\chi\right)	\right]
	+ \frac{1}{6}\frac{\xi^2\eta^{\prime 2}}{\eta^2\chi^2}
	+ \frac{1}{6}\frac{\xi^{\prime 2}}{\chi^2}
	+ \frac{2}{3}\frac{\phi^{\prime 2}}{\phi^2}
\nonumber\\
	& + & \left(
		\frac{4}{3}\frac{\phi^\prime}{\phi} - \frac{1}{3}\frac{\xi\xi^\prime}{\chi^2}
	\right)\frac{\eta^\prime}{\eta}
	+ \left(\frac{4}{3}\frac{\phi^\prime}{\phi} + \frac{2}{3}\frac{\eta}{\eta^2}\right)
	\frac{\chi^\prime}{\chi}
\nonumber
\end{eqnarray}
behaves asymptotically as $R\to 10/3 \Lambda$; for $\Lambda>0$ under consideration, this corresponds to an asymptotically anti-de Sitter spacetime.

Particularly simple is the case of a linear spinor field  (i.e., when $\lambda=0$). For this case, Eq.~\eqref{chi_eq} has the following particular solution:
$$
\chi=\chi_0\sqrt{\cosh\left(\sqrt{\frac{2\Lambda}{3}} r\right)}.
$$
Substituting this in the Dirac equations~\eqref{DE1} and \eqref{DE2} (which are now separated for the functions $A$ and $B$)
and taking into account the need to satisfy the constraint equation~\eqref{En_constr}, one has the following
 solutions for the spinor fields (asymmetric with respect to the domain wall):
$$
A=A_0 e^{m r}\sech{\left(\sqrt{\frac{2\Lambda}{3}} r\right)}, \quad
B=\frac{\Lambda}{2 A_0 m}e^{-m r}\sech{\left(\sqrt{\frac{2\Lambda}{3}} r\right)}.
$$
Asymptotically, on both sides of the domain wall, there is the following behavior of the spinor fields:
$$
\text{as}\quad r\to \pm \infty: A\to 2 A_0 \exp{\left[\mp\left(\sqrt{\frac{2\Lambda}{3}}\mp m\right)r\right]},
\quad B\to \frac{\Lambda}{ A_0 m} \exp{\left[\mp\left(\sqrt{\frac{2\Lambda}{3}}\pm m\right)r\right]} .
$$
Hence we see that regular solutions for the spinor fields are possible only if
$$
\sqrt{\frac{2\Lambda}{3}}-m>0.
$$

\subsection{The case of nonzero $\Omega$ and $M$}

\begin{figure}[t]
\includegraphics[width=1\linewidth]{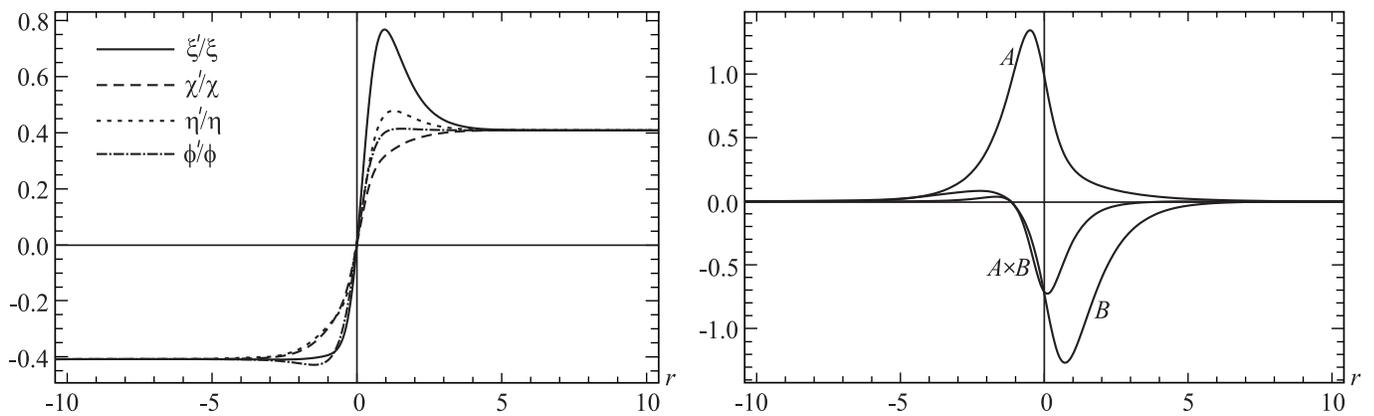}
\caption{Examples of the solutions with $\Lambda=1$, $\Omega=0.1$, $m=0.2$, and $M=0.05$ for the nonlinear spinor field with the nonlinearity parameter $\lambda=-1$.
}
\label{fig_asym_sol}
\end{figure}

When the free parameters $\Omega$ and $M$ are nonzero, solutions for the metric functions are not already symmetric with respect to the domain wall.
In this case regular solutions do exist both for simultaneously nonzero $\Omega$ and $M$ and in the case when one of these parameters is zero.
Using the boundary conditions~\eqref{BC_metr} and \eqref{BC_spinor}, the numerical solutions to the equations~\eqref{En1}-\eqref{DE2}
are exemplified in Fig.~\ref{fig_asym_sol}.

\begin{figure}[t]
\includegraphics[width=1\linewidth]{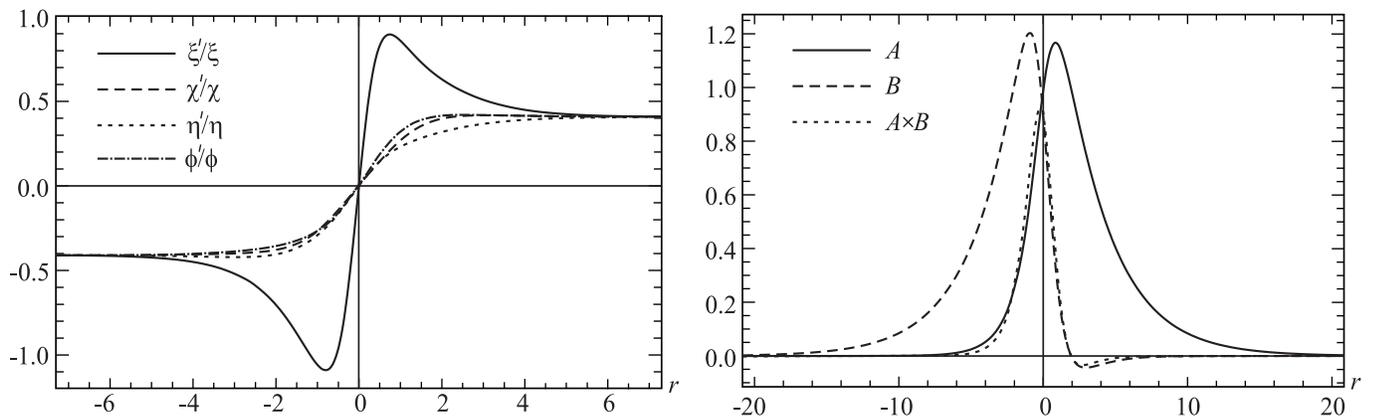}
\caption{Examples of the solutions with
  $\Lambda=1$, $\Omega=0.1$, $m=0.5$, and $M=0.05$ for the linear spinor field.
  }
\label{fig_asym_sol_lambda_0}
\end{figure}

In the case of a linear spinor field (i.e., when $\lambda=0$), there are also regular solutions shown in Fig.~\ref{fig_asym_sol_lambda_0}.

It is seen from the graphs given in Figs.~\ref{fig_asym_sol} and~\ref{fig_asym_sol_lambda_0}
that, as in the case of  $\Omega=M=0$ considered above, we deal with an asymptotically anti-de Sitter spacetime.

\subsection{The case without spinor field
}

In the absence of the spinor field (but when $\Lambda$ is present) there are no regular solutions to Eqs.~\eqref{En1}-\eqref{DE2}.

\subsection{
Three-dimensional energy density of the system
}
In this subsection we consider a question of the energy per unit volume of the domain wall, which is defined as
\begin{align}
\label{energ_int}
	&
	E_{\text{3D}} = \int_{-\infty}^{\infty} T_0^0\sqrt{^5 g}dr \equiv 
	\frac{1}{2} \int_{-\infty}^{\infty} \frac{\phi^2 }{\chi}
	\Big\{
		B^2 \left[
			M \xi \left(\xi + \chi \right) - \Omega \eta \left(\xi + 2 \chi \right)
		\right]
\\
	&
	\hspace{4.9cm} + 
	A^2 \left[
		M \xi \left(- \xi + \chi \right) - \Omega \eta \left(- \xi + 2 \Omega\chi  \right)
	\right] 
	+ 4 \lambda A^2 B^2 \eta \chi^2
	+ A B \eta \left(\xi\chi^\prime - \chi \xi^\prime\right)
	\Big\} dr .
\nonumber
\end{align}
It is evident that for a physically realistic system this quantity must be finite. Taking into account that asymptotically the metric functions diverge exponentially
(an anti-de Sitter spacetime) and the spinor functions decrease exponentially, it is necessary to follow the behavior of the expression in the integrand depending on the values
of the free parameters of the system. To do this, let us write down the corresponding asymptotic  (as $r\to \infty$) expressions for the metric functions and spinor fields:
\begin{align}
\label{asymp_metr}
&\xi\to \xi_\infty e^{\sqrt{\Lambda/6} r},\quad \chi\to \chi_\infty e^{\sqrt{\Lambda/6} r},\quad
\eta\to \eta_\infty e^{\sqrt{\Lambda/6} r},\quad \phi\to \phi_\infty e^{\sqrt{\Lambda/6} r},\\
&A\to C_1 e^{-\left(m+\sqrt{3\Lambda/2}\right) r}+C_2 e^{-\left(\sqrt{2\Lambda/3}-m\right) r}, \quad
B\to \tilde{C}_1  e^{-\left(m+\sqrt{2\Lambda/3}\right) r} ,
\label{asymp_spinor}
\end{align}
where $\xi_\infty, \chi_\infty, \eta_\infty, \phi_\infty$ are some constants,
$C_1$ and $C_2$ are integration constants and
\begin{equation}
\tilde{C}_1=C_1\frac{\left(12m+\sqrt{6\Lambda}\right)\eta_\infty \xi_\infty}{6\left[\Omega\eta_\infty-M\left(\xi_\infty+\chi_\infty\right)\right]} .
\label{C_1_expr}
\end{equation}
It is seen from Eq.~\eqref{asymp_spinor} that in the general case asymptotically decaying solution for the function $A$ can exist only when
\begin{equation}
0<m<\sqrt{\frac{2}{3}\Lambda}.
\label{rest_m}
\end{equation}

Next, substituting \eqref{asymp_metr} and \eqref{asymp_spinor} in  Eq.~\eqref{energ_int}, we have the following asymptotic  (as $r\to \infty$) form
for the expression in the integrand:
\begin{align}
\frac{1}{2}\frac{\phi_\infty^2}{\chi_\infty}&\Big\{
C_2^2\left[\Omega\eta_\infty \left(\xi_\infty-2\chi_\infty\right)+M\xi_\infty\left(\chi_\infty-\xi_\infty\right)
\right]e^{-\left(\sqrt{\Lambda/6}-2 m\right) r}\nonumber\\
&-\tilde{C}_1^2\left[\Omega\eta_\infty \left(\xi_\infty+2\chi_\infty\right)-M\xi_\infty\left(\chi_\infty+\xi_\infty\right)
\right]e^{-\left(\sqrt{\Lambda/6}+2 m\right) r}\Big\} .
\label{under_int}
\end{align}
(Similar expression can also be obtained for the case of $r\to -\infty$.)
Hence we see that to obtain an asymptotically decaying expression,  it is necessary that
$$
m<\frac{1}{2}\sqrt{\frac{\Lambda}{6}} .
$$
Since this condition is stronger than \eqref{rest_m}, the resulting restriction on the system parameters is therefore
$$
0<m<\frac{1}{2}\sqrt{\frac{\Lambda}{6}} .
$$
When this condition is fulfilled, the expression in the integrand of Eq.~\eqref{energ_int} will decrease asymptotically, and the corresponding integral will be finite.

One further remark may be made, to complete this subsection. The formulas obtained above are invalid for the case considered in Sec.~\ref{Omega_M_0}
where $\Omega=M=0$ [cf. Eqs.~\eqref{C_1_expr} and \eqref{under_int}]. However, proceeding in a similar manner, one can show that the energy of the configuration from
Sec.~\ref{Omega_M_0} is also finite.

\subsection{Angular momentum density of the domain wall
}

According to the definition of the angular momentum five-tensor,
$$
	M_{A B} = x_A T^t_{\phantom{t} B} - x_B T^t_{\phantom{t} A} ,
$$
one can introduce the angular momentum density three-vector
$$
	L_i = \frac{1}{2} \epsilon_{i j k} M^{j k}  ,
$$
as well as the ``second angular momentum density three-vector''
$$
	\mathcal{L}_i = M_{i 5} .
$$
Using the components of the energy-momentum tensor \eqref{A_10}-\eqref{A_60} given in the Appendix~\ref{app_A},
one can find the following expressions for the components of the angular momentum density $L_{x, y}$ and  the
``second angular momentum density''~$\mathcal{L}_z$:
\begin{align}
	L_{x,y} = & \begin{pmatrix}
		-y	\\
		x
	\end{pmatrix} T^t_{\phantom{t} z} ,
\label{3_e_40}\\
	\mathcal{L}_z = & - r T^t_{\phantom{t} z} .
\label{3_e_50}
\end{align}
In cylindrical coordinates,  Eq.~\eqref{3_e_40} yields
$$
	L_{\varphi} = \rho T^t_{\phantom{t} z} \quad \text{with}\quad \rho = \sqrt{x^2 + y^2} .
$$

\begin{figure}[t]
\includegraphics[width=.5\linewidth]{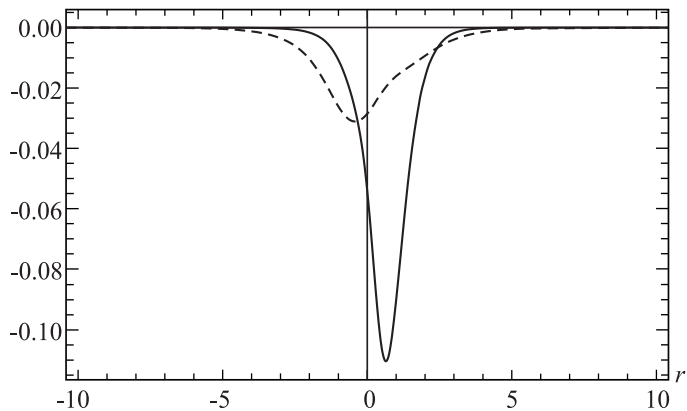}
\caption{The distributions of the component $T^t_{\,\, z}$ for the solutions represented in Fig.~\ref{fig_asym_sol} (for the nonlinear spinor field; shown by the solid line)
and in Fig.~\ref{fig_asym_sol_lambda_0} (for the linear spinor field; shown by the dashed line).
  }
\label{fig_Ttz}
\end{figure}

The typical spatial distributions of the component $T^t_{\phantom{t} z}$ [which is given by the expression~\eqref{A_20}]
appearing in Eqs.~\eqref{3_e_40} and \eqref{3_e_50} are plotted in Fig.~\ref{fig_Ttz}.
Notice that for the case considered in Sec.~\ref{Omega_M_0} where  $\Omega=M=0$
and the metric functions are linearly dependent, the expression~\eqref{A_20} is identically zero.

The physical reason for the appearance of the angular momenta $L_{\varphi}$ and $\mathcal{L}_z$
is the existence of the current density $j^A=\bar \psi \gamma^A \psi$, which has the following nonvanishing componets:
$$
	j^t =  \frac{1}{\chi}\left(A^2 + B^2\right) , \quad
	j^z =  \frac{A^2}{\eta} \left(
	1 - \frac{\xi }{\chi }
	\right)
	- \frac{B^2 }{\eta}\left(
	1+\frac{\xi }{\chi }
	\right) .
$$

\section{Spin of test fermions on the domain wall
}

In this section we consider the properties of spin operators of test fermions confined on the domain wall.
In order to distinguish these fermions from the spinor field $\psi$ supporting the domain wall, we denote them as $\chi$.
(Note that in this section we use the letters $\chi, \eta$, and $\phi$ to denote the spinor functions.
We have used above the same symbols for the metric functions, but this should not lead to misunderstanding.)
The consideration given below can also be applied to any domain wall, not necessarily to the domain walls obtained above.

In a four-dimensional spacetime, three-dimensional spin operators can be represented by one three-vector  $\Sigma_i$ dual to the
$i, j = 1,2,3$ components of the generators of the Lorentz algebra,
$$
	\Sigma_i = \frac{1}{2}\epsilon_{i j k} \sigma_{j k} =
	\frac{\imath }{4} \epsilon_{i j k} \left[ \gamma_j , \gamma_k	\right] .
$$
This construction can be adopted to a five-dimensional spacetime where the domain wall under consideration is embedded.
In the five-dimensional spacetime, the generators of the Lorentz algebra are
$$
	\Sigma^{a b} =\frac{\imath}{4} \left[ \gamma^a , \gamma^b	\right] \quad \text{with}\quad  a, b = 0,1, 2,3, 5.
$$
 Generalising the construction of the spin in a four-dimensional spacetime, spin operators in a five-dimensional spacetime will be the matrices
$$
	\Sigma^{\alpha \beta} =\frac{\imath}{4} \left[ \gamma^\alpha , \gamma^\beta	\right] \quad \text{with}\quad
	\alpha, \beta = 1,2,3,5 .
$$

Unlike the four-dimensional spacetime, it looks impossible to dualize these matrices;
for this reason, the spin operators form now not a vector, but a spatial tensor  $\Sigma^{\alpha \beta}$.
Being antisymmetric, this tensor is equivalent to two three-vectors
$$
	\Sigma_i = \frac{1}{2} \epsilon_{i j k} \Sigma_{j k} , \quad \mathcal{S}_i =  \Sigma_{5 i}
$$
with $i, j , k = 1, 2, 3$.
The operators  $\Sigma_i$ have the same form as the standard spin operators in a four-dimensional spacetime.
In turn, the operators $\mathcal{S}_i$ have the form
$$
	\mathcal{S}_i = \frac{1}{2}
	\begin{pmatrix}
		0														&	\sigma_{i}	\\
		\sigma_{i}  &	0
	\end{pmatrix} .
$$
The commutation relations for these operators are
\begin{align}
	\left[ \Sigma_i, \Sigma_j\right] = &\imath \epsilon_{i j k} \Sigma_k ,
\label{4_90}\\
	\left[ \mathcal{S}_i, \mathcal{S}_j\right] = &\imath \epsilon_{i j k} \Sigma_k ,
\label{4_100}\\
	\left[ \Sigma_i, \mathcal{S}_j\right] = &\imath \epsilon_{i j k} \mathcal{S}_k .
\label{4_110}
\end{align}
The operators $\Sigma_i$ are standard operators of the projection of the spin on the  $x^i$-axis.
Let us call the operators $\mathcal{S}_i$ as the operators of the projection of the ``second spin'' on the  $x^i$-axis.
Consistent with the commutation relations \eqref{4_100} and \eqref{4_110}, it is seen that the projections of the ``second spin'' on the  $x^i$ and $x^j$  axes  (with $i \neq j$)
cannot be measured simultaneously. Also, it is seen from Eq.~\eqref{4_110} that eigenvalues of the operators
 $\Sigma_i$ and $\mathcal{S}_j$ can be measured simultaneously only if  $i = j$.

The eigenvalue problem for the operator of the ``second spin'' is
$$
	\mathcal{S}_i \chi_{i, n,\pm} = s_{i, \pm} \chi_{i, n,\pm} ,
$$
where $s_{i, \pm}$ are eigenvalues and  $\chi_{i, n,\pm}$  are eigenspinors of the operator of the ``second spin'',
and the index $i$ describes the components of the ``second spin'' along the  $x,y,z$ axes, the index
$n$ corresponds to the number of the eigenspinor, and  $\pm$ corresponds to the projection of the ``second spin''  on the
 $i$-axis.
Solving this problem, we get the following eigenvalues and eigenspinors:
\begin{align}
	s_{x,-} = & - \frac{1}{2}, \chi^T_{x, 1,-} = \left\lbrace -1, 0, 0, 1\right\rbrace,
	\chi^T_{x, 2,-} = \left\lbrace 0, -1, 1, 0\right\rbrace ,
\label{4_130}\\
	s_{x,+} = & + \frac{1}{2}, \chi^T_{x, 1,+} = \left\lbrace 1, 0, 0, 1\right\rbrace,
	\chi^T_{x, 2,+} = \left\lbrace 0, 1, 1, 0\right\rbrace ,
\label{4_140}\\
	s_{y,-} = & - \frac{1}{2}, \chi^T_{y, 1,-} = \left\lbrace \imath, 0, 0, 1\right\rbrace,
	\chi^T_{y, 2,-} = \left\lbrace 0, - \imath, 1, 0\right\rbrace ,
\label{4_150}\\
	s_{y,+} = & + \frac{1}{2}, \chi^T_{y, 1,+} = \left\lbrace - \imath, 0, 0, 1\right\rbrace,
	\chi^T_{y, 2,+} = \left\lbrace 0, \imath, 1, 0\right\rbrace ,
\label{4_160}\\
	s_{z,-} = & - \frac{1}{2}, \chi^T_{z, 1,-} = \left\lbrace 0, 1, 0, 1\right\rbrace,
	\chi^T_{z, 2,-} = \left\lbrace -1, 0, 1, 0\right\rbrace ,
\label{4_170}\\
	s_{z,+} = & + \frac{1}{2}, \chi^T_{z, 1,+} = \left\lbrace 0, -1, 0, 1\right\rbrace,
	\chi^T_{z, 2,+} = \left\lbrace 1, 0, 1, 0\right\rbrace .
\label{4_180}
\end{align}
Note that the eigenspinors are orthogonal each other (for simplicity, we omitted the indices $i$ and $\pm$):
$$
	 \bar \chi_1 \chi_2 = 0 .
$$

\subsection{Eigenspinors of the operator of the ``second spin'' and the Bell states}

In this subsection we show that some eigenspinors of the operator of the ``second spin'' \eqref{4_130}-\eqref{4_160} are the Bell states.
To do this, let us introduce quantum states $\left|  0 \right\rangle $ and $\left| 1\right\rangle $
for some quantum system that can take only two possible states. For example, this can be the projection of the spin on the $z$-axis.
These quantum states can be written in the form of the following Weyl spinors:
$$
	\left|  0 \right\rangle =
		\begin{pmatrix}
		0	\\
		1
	\end{pmatrix} ,
	\left| 1\right\rangle =
	\begin{pmatrix}
		1	\\
		0
\end{pmatrix} .
$$
Then, apart from a normalization factor, the eigenspinors \eqref{4_130}-\eqref{4_180} can be represented as
 \begin{align}
	\chi_{x, 1, \pm} = & 	\left|  0 \right\rangle \otimes 	\left|  0 \right\rangle
	\pm \left|  1 \right\rangle \otimes 	\left|  1 \right\rangle ,
\label{4_a_20}\\
	\chi_{x, 2, \pm} = & 	\left|  0 \right\rangle \otimes \left| 1 \right\rangle
	\pm \left|  1 \right\rangle \otimes \left| 0 \right\rangle ,
\label{4_a_30}\\
	\chi_{y, 1, \pm} = & 	\left|  0 \right\rangle \otimes \left| 0 \right\rangle
	\mp \imath \left|  1 \right\rangle \otimes \left| 1 \right\rangle ,
\label{4_a_40}\\
	\chi_{y, 2,\pm} = & 	\left|  0 \right\rangle \otimes \left| 1 \right\rangle
	\pm \imath \left|  1 \right\rangle \otimes \left| 0 \right\rangle ,
\label{4_a_50}\\
	\chi_{z, 1, \pm} = & \mp	\left| 1 \right\rangle \otimes \left| 0 \right\rangle
	+ \left|  0\right\rangle \otimes \left| 0 \right\rangle = \left(
		 \mp	\left| 1 \right\rangle + \left|  0\right\rangle
	\right) \otimes \left| 0 \right\rangle ,
\label{4_a_60}\\
	\chi_{z, 2, \pm} = & \pm	\left| 1 \right\rangle \otimes \left| 1 \right\rangle
	+ \left|  0\right\rangle \otimes \left| 1 \right\rangle =
	\left(
		 \pm	\left| 1 \right\rangle + \left|  0\right\rangle
	\right) \otimes \left| 1 \right\rangle .
\label{4_a_70}
\end{align}
The quantum states  \eqref{4_a_20}-\eqref{4_a_50} are called the Bell states or sometimes the EPR states or EPR pairs
(after Bell or Einstein, Podolsky, and Rosen,  who first pointed out the strange properties of such states).
Strictly speaking,  the entangled states~\eqref{4_a_20}-\eqref{4_a_50} differ from the Bell states by some factors on the right-hand sides of Eqs.~\eqref{4_a_20}-\eqref{4_a_50}.
In quantum calculations, the Bell states describe an entangled pair of qubits. The difference of the quantum states
\eqref{4_a_20}-\eqref{4_a_50} from those given by Eqs.~\eqref{4_a_60}-\eqref{4_a_70} is that the former cannot be represented as a tensor product of states, while the latter,
 as one sees from the right-hand sides of Eqs.~\eqref{4_a_60}-\eqref{4_a_70},
can be represented as  tensor products of some quantum states.

It must be mentioned here that all the quantum states of the spin $\Sigma_i$ can be represented as  tensor products:
\begin{align}
	\chi_{x, \pm} = & \left(
		\left| 0 \right\rangle + \left| 1 \right\rangle
	\right) \otimes \left(
	\left| 0 \right\rangle \pm \left| 1 \right\rangle
	\right) ,
\quad
	\chi_{y, \pm} =  \left(
		\left| 0 \right\rangle + \left| 1 \right\rangle
	\right) \otimes \left(
		\left| 0 \right\rangle \pm \imath \left| 1 \right\rangle
	\right) ,
\nonumber\\
	\chi_{y, +} = & \left(
	\left| 0 \right\rangle + \left| 1 \right\rangle
		\right) \otimes \left| 1 \right\rangle ,
\hspace{1.5cm}
		\chi_{y, -} =  \left(
	\left| 0 \right\rangle + \left| 1 \right\rangle
	\right) \otimes \left| 0 \right\rangle .\nonumber
\end{align}

\subsection{Nonrelativistic limit for the test fermions
}

In order to understand what extra features (compared with the four-dimensional case) appear for the test fermions living on the domain wall
embedded in the five-dimensional  bulk, we consider the nonrelativistic limit of the five-dimensional Dirac equation, i.e.,
the Pauli equation in  the five-dimensional  bulk spacetime. In such case we should rewrite the Dirac equation for a flat five-dimensional spacetime
and take into account the interaction with a gravitational field which is now described by the potential energy
 $m \Phi$, where $m$ is the mass of a particle and  $\Phi$ is the Newtonian gravitational potential.

In order to obtain the nonrelativistic limit for the five-dimensional Dirac equation (the Pauli equation), we proceed in the standard way
and write the Dirac equation in the following form  (for clarity, in this subsection we resurrect $\hbar$ and $c$ in the equations):
\begin{equation}
	\imath \hbar \partial_t \begin{pmatrix}
		\phi	\\
		\eta
	\end{pmatrix} = c \begin{pmatrix}
		\sigma_i \hat \Pi_i	\eta \\
		\sigma_i \hat \Pi_i \phi
	\end{pmatrix} + \imath c \hat \Pi_5
	\begin{pmatrix}
		\eta	\\
		- \phi
	\end{pmatrix} +
	m c^2 	\begin{pmatrix}
		\phi	\\
		- \eta
	\end{pmatrix} + \left( e A_0 + m \Phi \right)
	\begin{pmatrix}
		\phi	\\
		\eta
	\end{pmatrix} ,
	\label{4_b_30}
\end{equation}
where $\hat \Pi_i = \hat p_i - e/c A_i$,  $\hat p_i$ being the three-dimensional momentum operator,
$\hat \Pi_5=\hat p_5 - e/c A_5$, and
$A_B = \left(A_0, A_i, A_5 \right)$ is the electromagnetic five-potential. Also, as in the four-dimensional case,
the four-component spinor $\chi$ is again decomposed into two two-component spinors,
$$
	\chi = 	\begin{pmatrix}
		\phi	\\
		\eta
	\end{pmatrix} .
$$
The relativistic energy of a particle described by Eq.~\eqref{4_b_30} contains also its rest energy $m c^2$.
In arriving at the nonrelativistic approximation, this energy must be excluded by introducing a new function $\chi\to \chi e^{-i m c^2 t/\hbar}$.
Next, using the same nonrelativistic approximations as those employed in the four-dimensional case,
$$
	\left| \imath \hbar \dot \eta \right| ,
	\left| e A_0 \eta \right|, \text{ and } \left| m \Phi \eta \right|
	\ll \left| m c^2 \eta \right| ,
$$
one can obtain from the lower component of Eq.~\eqref{4_b_30} the relation
$$
	\eta = \frac{\sigma_i \hat \Pi_i - \imath \hat{\Pi}_5 }{2 m c} \phi .
$$
In this way, after substituting of this into the upper part of the Eq.~\eqref{4_b_30},
we finally arrive at the required Pauli equation for the five-dimensional spacetime:
\begin{equation}
	\imath \hbar \dot{\phi} =
	\left[
		\frac{1}{2 m} \left(
			\hat{\vec p} - \frac{e}{c} \vec A
		\right)^2 + \frac{1}{2 m} \left(
			\hat{p}_5 - \frac{e}{c} A_5
		\right)^2 - \frac{e \hbar}{2 m c} \vec \sigma \left(
			\vec H + \vec{\mathcal{H}}
		\right) + \left( e A_0 + m \Phi \right)
	\right] \phi ,
\label{4_b_70}
\end{equation}
where the vector $\vec H\equiv H_i = (1/2) \epsilon_{i j k} F^{jk}$ is the magnetic field and  the vector $\vec{\mathcal{H}}\equiv \mathcal{H}_i = F_{i 5}$
is the ``second magnetic field''.

The physical meaning of the new terms appearing in Eq.~\eqref{4_b_70} (compared with the four-dimensional Pauli equation) is as follows.
Because of the presence of the second term in the square brackets, there are quantum corrections to the motion of a test fermion which is located on the domain wall embedded in the  bulk.
These corrections are not associated with the spin of the fermion.
In turn, the presence of the term with
 $\vec{\mathcal{H}}$ results in quantum corrections to the motion of the particle which are due to the presence of the spin.
 This term describes the potential energy of a particle with a spin in the ``second magnetic field.''
Hence, if the gradient of this field differs from zero,  trajectories of particles possessing different projections of a spin on some
direction in the presence of the ``second magnetic field'' will be different from one another (the Stern-Gerlach effect).

Thus, if the motion of a test fermion takes place on the domain wall, only  the ``second magnetic field'' $\vec{\mathcal{H}}$
in the five-dimensional Pauli equation~\eqref{4_b_70} will lead to the corrections associated with the fact that our world is embedded as a domain wall
in the five-dimensional spacetime.
\textit{This correction allows the possibility of an experimental verification of the existence of extra dimensions. }

\section{Conclusion}

In the first part of the present paper, we have considered the self-consistent set of the Einstein-Dirac equations and shown that
there exist plane symmetric solutions for one gravitating classical spinor field. These solutions
can be treated as describing a thick domain wall embedded in a five-dimensional spacetime.

The second part of the paper is devoted to exploring the properties and behavior of test fermions localized on \textit{any} domain wall. In a five-dimensional spacetime, we cannot introduce the concept of a spin vector since the procedure of dualization of the spatial part of the spin tensor results in a tensor, and not in a vector. Nevertheless, the spatial part of the five-dimensional spin tensor can be represented as some combination of two three-vectors. The first of these vectors corresponds to an ordinary spin, and we referred to the second one as the ``second spin''. It was shown that eigenvalues of the operator of the ``second spin'' are also $\hbar/2$, and some eigenspinors are the Bell states.

To study the behavior of test fermions, we have obtained a nonrelativistic approximation for the five-dimensional Dirac equation,
which is a natural generalization of the Pauli equation containing now additional terms that describe
(i)~the presence of the component $A_5$  of the electromagnetic potential; and (ii)~the interaction of a spin-$1/2$
particle with the components $F_{i 5}$ of the electromagnetic field five-tensor, which we called the ``second magnetic field.''
The presence of the term with the component $A_5$ of the electromagnetic potential must lead to the appearance of the Aharonov-Bohm effect;
however, to register this effect, it is necessary to escape beyond the domain wall, i.e., one has to have a solenoid directed along the fifth dimension.
By contrast, the effects associated with the presence of the ``second magnetic field'' can be  measured directly, since this ``second magnetic field''
interacts with the spin directly.

Summarizing the results obtained,
\begin{itemize}
	\item We have found regular symmetric and asymmetric thick domain wall solutions supported by one classical spinor field. In doing so,
we have examined both linear and nonlinear spinor fields by choosing different values of the free parameters of the system.
Note that, as the numerical calculations indicate, for the regular solutions to exist, the presence of the cosmological constant is necessary.
	\item We have studied the properties of these domain wall solutions which depend on the presence or absence of the nonlinearity and mass term of the spinor field.
It is shown that the nonlinearity of the spinor field brings no qualitative changes to the behavior of the solutions,
although the numerical values of the fields are quite different.
    \item We have demonstrated that there are the nonvanishing components  $L_\varphi$ and $\mathcal{L}_z$ of the angular momenta density
    associated with the current of the spinor field. The presence of the component $L_\varphi$ allows the possibility of an experimental verification of the existence of the domain wall
    by using the spin-orbit interaction.
    \item The behavior and properties of \textit{test} fermions living on \textit{any} domain wall have been investigated:
	\begin{itemize}
		\item It has been shown that the spatial part of the five-dimensional spin tensor can be represented as a combination of two vectors,
the first of which is an ordinary spin, and the second vector can be called the ``second spin.''
		\item The eigenvalues ($\hbar/2$) and eigenspinors for the ``second spin'' have been found. It has been demonstrated that some eigenspinors are the Bell states.
		\item The nonrelativistic approximation for the five-dimensional Dirac equation (the Pauli equation) has been derived.
		\item It has been shown that the Pauli equation obtained contains extra terms associated with the fact that the domain wall is embedded in a five-dimensional bulk.
One of these terms is related to the appearance of the component $A_5$ in the Pauli equation, and the second one is due to the fact that the spin interacts with
the components $F_{i 5}$ of the electromagnetic field five-tensor (one can call it the ``second magnetic field'').
 		\item It has been demonstrated that the interaction of the ``second magnetic field'' with the spin results in observational effects which enable one to verify experimentally the hypothesis of
 the existence of extra dimensions.
 	\end{itemize}
\end{itemize}

\section*{Acknowledgments}

This research has been funded by the Science Committee of the Ministry of Science and Higher Education of the Republic of Kazakhstan
(Grant  No.~AP14869140, ``The study of QCD effects in non-QCD theories'').
We are also grateful to V.~Ivashchuk for fruitful discussions.

\appendix

\section{The components of the energy-momentum tensor}
\label{app_A}

For the metric \eqref{metr} and
$\mathfrak{Ansatz}$ \eqref{2-110}, the mixed  energy-momentum tensor~\eqref{2-40} yields the following nonzero components:
\begin{align}
\label{A_10}
	T^t_{\phantom{t} t} = & 2 \lambda A^2 B^2
\nonumber\\
	&+ \frac{A^2}{2 \chi} \left[
		 \Omega \left(\frac{ \xi}{\chi} - 2 \right)
		 + M \left(\frac{\xi}{\eta} - \frac{\xi^2}{\eta \chi} \right)
	\right] +
	\frac{A B}{2} \left(
		\frac{\xi  \chi '}{\chi ^2} - \frac{\xi '}{\chi }
	\right)
	+ \frac{B^2}{2 \chi} \left[
		M \left(  \frac{\xi}{\eta} + \frac{\xi^2}{\eta \chi} \right)
		- \Omega \left(\frac{\xi}{\chi} + 2\right)
	\right] ,
\\
	T^t_{\phantom{t} z} = &
	\frac{A^2}{2 \chi} \left[
		\Omega \frac{\eta}{\chi} - M \left(1+\frac{\xi}{\chi}  \right)
	\right]
	+ A B \frac{\eta}{2 \chi} \left(
		\frac{\chi '}{\chi} - \frac{\eta'}{\eta}
	\right)
	- \frac{B^2}{2 \chi} \left[
		 \Omega \frac{\eta}{\chi} + M \left(1- \frac{\xi}{\chi}   \right)
	\right] ,
\label{A_20}\\
	T^x_{\phantom{x} x} = & T^y_{\phantom{y} y} =
	2 \lambda A^2 B^2 ,
\label{A_30}\\
	T^z_{\phantom{z} t} = &
	A^2 \frac{ (\xi -\chi )^2}{2 \eta^2 \chi}
	\left[
		M \left(1+\frac{\xi}{\chi}  \right) - \Omega \frac{ \eta}{\chi}
	\right]
	+ A B \frac{\xi}{2 \eta \chi}
	\left[	 \eta' \left( \frac{\chi^2}{\eta \xi} - \frac{\xi}{\eta}\right)
        -\chi^\prime \left(  \frac{\xi}{\chi} + \frac{\chi}{\xi} \right)
		+ 2 \xi^\prime
	\right]
\nonumber \\
	+ & B^2 \frac{ (\xi + \chi )^2}{2 \eta^2 \chi}
	\left[
	M \left(1- \frac{\xi}{\chi}  \right) + \Omega \frac{ \eta}{\chi}
	\right] ,
\label{A_40}\\
\label{A_50}
	T^z_{\phantom{z} z} = & 2 \lambda A^2 B^2
\nonumber\\	
&+ \frac{A^2}{2 \eta } \left[
		M \left( \frac{\xi^2}{\chi^2} +\frac{\xi}{\chi} - 2 \right)
		- \Omega \frac{\eta\xi}{\chi^2}
	\right]
	+ A B \frac{\xi}{2 \chi} \left(
		\frac{\xi'}{\xi} - \frac{\chi'}{\chi}
	\right)
	+ \frac{B^2}{2 \eta } \left[
		M \left( - \frac{\xi^2}{\chi^2} +\frac{\xi}{\chi} + 2 \right)
		+ \Omega \frac{\eta\xi}{\chi^2}
	\right] ,
\\
	T^r_{\phantom{r} r} = &
	2 \lambda A^2 B^2 + B A' - A B'
	+ A B \frac{\xi}{2 \chi} \left(
		\frac{\eta '}{\eta } - \frac{\xi '}{\xi }
	\right) .
\label{A_60}
\end{align}

\end{document}